\documentclass[12pt]{article}

\usepackage{a4wide,graphics,graphicx,amsmath,amssymb,cite,nicefrac}
\usepackage[verbose]{wrapfig}
\usepackage{authblk,hyperref}
\catcode`@=11 \@addtoreset{equation}{section} \catcode`@=12

\begin{document}
\date{}
\title{
{\baselineskip -.2in
%\vbox{\small\hskip 4in \hbox{hep-th/07-10~~~~~~~~~~}}
%\vbox{\small\hskip 4in \hbox{IITM/PH/TH/2014/2}}
} 
\vskip .4in
\vbox{
{\bf \LARGE New Branches of Non-supersymmetric Attractors in $N=2$ Supergravity}
}}

\author{Prasanta K. Tripathy\thanks{email: prasanta@iitm.ac.in}}
\affil{\normalsize\it Department of Physics, \authorcr \it Indian Institute of Technology Madras, \authorcr \it Chennai 600036, India.}
\maketitle
\begin{abstract}

In this paper we analyse non-supersymmetric single centred extremal black hole solutions in $N=2$ supergravity theory coupled 
to $n$ vector multiplets with  purely cubic pre-potential in four dimensions. We consider the algebraic attractor equations in their 
most general form at the black hole horizon. We explicitly construct a new class of solutions for these attractor equations. These 
solutions are characterised by a set of involutory matrices. These involutions are obtained from a constraint involving the 
parameters in the pre-potential and generate new attractor points in the moduli space.
\end{abstract}

\newpage
\section{Introduction}

The attractor mechanism plays a central role in understanding the macroscopic origin of black hole entropy in gravity theories 
coupled to scalar fields \cite{Ferrara:2008hwa}. The mechanism shows that the scalar fields must run into a fixed point at the 
horizon irrespective of the value they take at the asymptotic infinity, with their values at the fixed point being entirely determined 
by the black hole charges. This explains why the black hole entropy is only a property of its horizon and must be independent 
of the asymptotic data involving the scalar fields.

The attractor mechanism has first been realised for supersymmetry preserving black holes in the context of four dimensional 
$N=2$ supergravity coupled with arbitrary number of  vector multiplets \cite{Ferrara:1995ih}. It has been promptly generalised 
for dyonic black holes \cite{Strominger:1996kf}. Various aspects of the mechanism have been studied on subsequently. One 
issue of great importance pertaining to the attractor mechanism is the existence of non-supersymmetric attractors 
\cite{Ferrara:1997tw,Goldstein:2005hq}. It has been shown afterwards that the attractor mechanism is really a 
consequence of extramility of the black holes. An extremal black hole may or may not be supersymmetric, however it always 
exhibits the attractor behaviour. 

One of the reasons the attractor mechanism plays an important role in understanding the black hole entropy is the uniqueness
of these attractors \cite{Wijnholt:1999vk}. Though a given single centred charge configuration appears to admit a unique 
supersymmetric attractor, it is not always the case. For example, the five dimensional supergravity admits multiple basin of 
supersymmetric attractors \cite{Kallosh:1999mz}. This, of course, depends on the topology of the moduli space of scalar fields 
coupled to the gravity multiplet \cite{Kallosh:1999mb}. For supersymmetric attractors in four dimensions, a classification of 
charged orbits has been carried out when the moduli space is a symmetric space \cite{Bellucci:2006xz}. More recently the 
uniqueness issue of supersymmetric attractors carrying $D0-D4-D6$ charges has been investigated in detail \cite{Manda:2015zoa}. 
A classification of all the supersymmetric solutions has been carried out for the above charge configuration. For this class of 
black holes it has been shown that there exists domains in the charge lattice such that the attractor solution is unique in a 
given domain. The moduli space metric becomes degenerate at the boundaries of these domains and hence single centred
black hole ceases to exist at these boundaries. However the black hole undergoes a kind of phase transition as one changes 
the values of the charges from one domain to the other. The functional form of the attractor point as well as the entropy changes 
as well.

The non-supersymmetric attractors are very much similar to their supersymmetric counterparts in a number of aspects
\cite{Tripathy:2005qp}. For example, the functional form of the respective entropies are identical for a given charge configuration. 
In the case of  axion free attractors, there is a set of first order flow equations determining the exact behaviour of the black hole in 
space-time \cite{Ceresole:2007wx,LopesCardoso:2007qid}. These equations are obtained upon the extermization of a fake 
superpotential which is analogous to the central charge for supersymmetric black holes. Thus it is worth asking if there exist 
analogous results in the case non-supersymmetric attractors. 

The goal of the paper is to explore these new branches in non-supersymmetric extremal black holes. In the next section we will
review the required background to study these solutions. In \S3 we will briefly outline the previously known extremal 
configurations. Subsequently in \S4 we will analyse the attractor equations and will solve them with a specific ansatz. Finally, 
we will summarise our results in \S5.

\section{The Model}

In the present work we will focus $N=2$ supergravity theory in four dimensions coupled to $n$ abelian vector multiplets. The 
bosonic part of the Lagrangian density is given by  
\begin{eqnarray}\label{sugra}
{\cal L} = - \frac{R}{2} +  g_{a\bar b} \partial_\mu x^a \partial_\nu \bar x^{\bar b} h^{\mu\nu} 
-  \mu_{\Lambda\Sigma} {\cal F}^\Lambda_{\mu\nu}{\cal F}^\Sigma_{\lambda\rho} 
h^{\mu\lambda}h^{\nu\rho} -   \nu_{\Lambda\Sigma} {\cal F}^\Lambda_{\mu\nu}
*{\cal F}^\Sigma_{\lambda\rho} h^{\mu\lambda}h^{\nu\rho} \ .
\end{eqnarray}
We use the standard notations and conventions as in \cite{Ferrara:1997tw} to describe the system. In particular, we use 
$h_{\mu\nu}$ to denote the four dimensional space-time metric with $R$ being the corresponding Ricci scalar. The complex 
scalars $x^a$ parametrise the moduli space for $n$ scalar fields in the vector multiplet and $g_{a\bar b}$ is the metric on it. 
${\cal F}^\Sigma_{\lambda\rho}$ is the field strength for the gauge fields ${\cal A}^\Sigma_\mu$. The indices $\Lambda, \Sigma$
take $n+1$ values due to the presence of an additional gauge field coming from the gravity multiplet. The gauge couplings 
$ \mu_{\Lambda\Sigma}$ and $ \nu_{\Lambda\Sigma}$ are derived from the $N=2$ pre-potential $F$. In this paper we will
entirely focus on the purely cubic pre-potential.

 For static, spherically symmetric configurations carrying dyonic charges $(p^\Lambda, q_\Sigma)$ the system reduces to an 
 effective one dimensional theory with the Lagrangian density:
 \begin{equation}\label{leff}
 \mathcal{L}(U, x^a(\tau),\bar{x}^a(\tau))=\bigg(\frac{dU}{d\tau}\bigg)^2+g_{a\bar b} \frac{dx^a}{d\tau} \frac{d\bar x^b}{d\tau}
 +e^{2U} V_{\rm eff} \ , %(|Z|^2+|D_aZ|^2) \ ,
\end{equation}
with the corresponding Hamiltonian density being constrained to vanish  \cite{Ferrara:1997tw}. Here $U$ is the warp factor
appearing in the space-time metric:
\begin{equation}
 ds^2=e^{2U(\tau)}dt^2-e^{-2U(\tau)}(d\vec{x})^2 \ , 
\end{equation}
 and $\tau$ is the inverse of the radial separation $\tau = 1/r$. The effective black hole potential $V_{\rm eff}$ is determined 
 in terms of the K\"ahler potential $K$ and the superpotential $W$ which in turn are derived from the pre-potential $F$ as:
 \begin{equation}
%\label{kpsg}
K=-\ln \Big(i\sum_{\Lambda=0}^n \big[\bar X^{\Lambda}\partial_\Lambda F(X) - X^{\Lambda}\bar\partial_\Lambda \bar F(X)\big]\Big) \ ,
\end{equation}
and % the superpotential by,
\begin{equation}
\label{superpotsg}
W=\sum_{\Lambda=0}^n \big(q_\Lambda X^\Lambda - p^\Lambda \partial_\Lambda F\big) .
\end{equation}
The superpotential $W$ is related to the central charge $Z$ by $Z=e^{K/2}W$. Note that the physical scalar fields 
$x^a\ (a = 1,\ldots, n)$ appearing in the effective one dimensional Lagrangian \eqref{leff}  as well as in the supergravity 
Lagrangian \eqref{sugra} are given in terms of the symplectic sections $X^\Lambda$ as $x^a = \nicefrac{X^a}{X^0}$. 
The effective potential $V_{\rm eff}$ has the expression \cite{Ferrara:1997tw}: 
 \begin{equation}
  \label{epot}
  V_{\rm eff}=e^K\big[g^{a \bar b}\nabla_a W (\nabla_b W)^* + |W|^2\big],
\end{equation}
 where the action of the K\"ahler covariant derivative on $W$ is given by  $\nabla_a W\equiv \partial_aW+\partial_aK W$. 
 
 The supersymmetric attractors are obtained by extremising the central charge. The condition can be expressed in terms of 
 the superpotential $W$ as 
 \begin{equation} \label{susycond} \nabla_a W = 0 \ . \end{equation}
 The supergravity theory however admits more general black hole configurations. For extremal black holes, existence of a 
 regular horizon requires that the effective potential $V_{\rm eff}$  is extremized on it. For the effective potential \eqref{epot} 
 this condition can explicitly be stated as \cite{Tripathy:2005qp}:
 \begin{equation}
  g^{b\bar c} \nabla_a\nabla_b W \overline{\nabla_cW} + 2 \nabla_a W \overline{W}
+ \partial_a g^{b\bar c} \nabla_b W\overline{\nabla_c W} = 0~.
\label{extpot}
\end{equation}
Clearly, the supersymmetric configurations do satisfy the above equation. However there is a possibility of existing more 
general configurations which solve \eqref{extpot} for which $\nabla_a W \neq 0$. Such non-supersymmetric extremal black 
hole attractors have been explored extensively during the past decade and their properties have been studied in detail. In
the remaining part of this paper we will examine the equations of motion \eqref{extpot} more carefully and find some new 
solutions which were previously unknown.

 \section{Extremal Solutions}
 
 In the present work we will entirely focus on $N=2$ supergravity theories with the purely cubic pre-potential:
 \begin{equation}
\label{prelv}
F=D_{abc}{X^aX^bX^c  \over X^0}
\end{equation}
The parameters $D_{abc}$ are totally symmetric and take arbitrary values in general. However this pre-potential takes  
a prominent role because of its appearance in large volume compactification of type $IIA$ string theory on a  Calabi-Yau 
manifold ${\cal M}$. In this case the parameters $D_{abc}$ are no longer arbitrary and are given in terms of the triple 
intersection numbers of ${\cal M}$:
\begin{equation}
\label{defdabc}
D_{abc}=\frac{1}{6}\int_{\cal M} \alpha_a\wedge \alpha_b \wedge \alpha_c \ ,
\end{equation}
 with $\{\alpha_a\}$ denoting a basis of the integral cohomology group $H^2({\cal M},\mathbb{Z})$.
 
 We will focus on configurations carrying $\{q_0, p^a, p^0\}$ charges. From the string theory point of view these will 
 correspond to $D0-D4-D6$ configurations carrying $q_0$ number of $D0$-branes, $p^a$ number of $D4$-branes 
 wrapping four cycles dual to $\alpha_a$ and $p^0$ number of $D6$ branes wrapping ${\cal M}$.
 
 For convenience we will set $x^a = \nicefrac{X^a}{X^0}$ and choose the gauge $X^0=1$. With this choice of the gauge,
 the K\"ahler potential $K$ and the superpotential $W$ for the above configuration have respectively the expressions
 \begin{equation}
 K = - \ln\big( - i D_{abc} (x^a - \bar x^a) (x^b - \bar x^b) (x^c - \bar x^c)\big) \ ,
 \end{equation}
 and
\begin{equation}
\label{spd6}
W=(q_0 -3 D_{ab}x^a x^b + p^0 D_{abc}x^ax^bx^c) \ .
 \end{equation}

This configuration admits the well known supersymmetry preserving solution \cite{Shmakova:1996nz},  
$$ x^a  =  p^a  t  $$ with
\begin{eqnarray}
\label{susyd6}
t = {1\over 2D}\left(-p^0q_0 \pm  i \sqrt{q_0(4D -(p^0)^2 q_0)}\right) \ .
\end{eqnarray}
Here we use the notation $D = D_{abc} p^ap^bp^c$. For the attractor solution to be non-singular, we require 
$ q_0(4D -(p^0)^2 q_0)>0$.

The attractor equation \eqref{extpot} however admits more general extremal solution. Existence of such non-supersymmetric 
attractors were first investigated in \cite{Tripathy:2005qp} upon setting the ansatz $x^a = p^a t $. The real and imaginary
parts of $t$ are given respectively by
\begin{eqnarray}
\label{t1d6}
t_1 = \left\{
\begin{matrix}
 \frac{2}{s} {\left(1+{p^0 \over s}\right)^{1/3}-\left(1-{p^0 \over s}\right)^{1/3}\over \left(1+{p^0 \over s}\right)^{4/3}
+\left(1-{p^0 \over s}\right)^{4/3}} &
|{s\over p^0}|>1 \cr
{2\over p^0} {\left(1-{s\over p^0}\right)^{1/3}+\left(1+{s\over p^0}\right)^{1/3}\over \left(1-{s\over p^0}\right)^{4/3}+
\left(1+{s \over p^0}\right)^{4/3}} &
|{s\over p^0}| < 1  \ , \cr
\end{matrix}
\right.
\end{eqnarray}
and  %$t_2$ by:
\begin{eqnarray}
\label{t2d6}
t_2 = \left\{\begin{matrix}
 {4 s \over (s^2-(p^0)^2)^{1/3} \left((s+p^0)^{4/3}+(s-p^0)^{4/3}\right)} &
|{s \over p^0}|>1 \cr
 {4 s \over ((p^0)^2-s^2)^{1/3} \left((|p^0|+s)^{4/3}+(|p^0|-s)^{4/3}\right)} &
|{s \over p^0}|<1 \ .  \cr
\end{matrix}\right.
\end{eqnarray}
Here we introduced the variable $s=\sqrt{(p^0)^2-{4D\over q_0}}$ for convenience. Note that the above non-supersymmetric
solution is non-singular provided $ q_0(4D -(p^0)^2 q_0) < 0$.
 
\section{New Branches}

More recently the supersymmetric conditions for black holes carrying $D0-D4-D6$ charges were analysed in more detail 
\cite{Manda:2015zoa}. It was realised that the configuration described by \eqref{susyd6} is not the most general solution 
for supersymmetric attractor carrying these charges. There exists family of solutions determined by  involutory matrices 
${I^a}_b$ satisfying 
\begin{eqnarray}
 D_{abc}{I^b}_e{I^c}_f = D_{aef}  \label{iabda} \ .
\end{eqnarray}
The most general solution for \eqref{susycond} is given by $x^a = x_1^a + i x_2^a$ with
\begin{eqnarray}
x_1^a &=& \frac{1}{p^0}\bigg(p^a - \frac{D-\frac{1}{2}q_0{p^0}^2}{D_c{I^c}_dp^d}{I^a}_bp^b\bigg)\ ,\label{eq:g2} \\
x_2^a &=& \frac{1}{p^0}\bigg(\ 1-{\bigg(\frac{D-\frac{1}{2}q_0{p^0}^2}{D_c{I^c}_dp^d}\bigg)}^2\ \bigg)^{1/2}{I^a}_bp^b\ . \label{eq:g1}
\end{eqnarray}
For all these solutions the charges must satisfy $q_0(4D -(p^0)^2 q_0)>0$. However, this is not the only criteria for the 
existence of a smooth solution. A more fundamental requirement is the positive definiteness of the moduli space metric 
at the attractor point. This requirement divides the charge lattice into several domains and different involutions gives rise
to unique attractor configurations in each such domains of the charge lattice \cite{Manda:2015zoa}.

In the following we will derive analogous solutions for the non-supersymmetric attractors. We will first obtain the equations
of motion in its general form for the pre-potential \eqref{prelv} and then analyse them to obtain specific solutions. Let us now
compute various terms in \eqref{extpot}. We denote $x^a = x^a_1 + i x^a_2$, and introduce the notations $D_{ab} = D_{abc} p^c, \
D_a = D_{ab} p^b,  \ \nu_{ab} = D_{abc}x_2^c, \ \nu_a = \nu_{ab}x_2^b, \nu = \nu_a x_2^a$  for convenience. We further 
introduce the variable $\omega^a = (\nicefrac{p^a}{p^0}) - x_1^a$  and define $\mu_{ab} = D_{abc}\omega^c, \ 
\mu_a = \mu_{ab}\omega^b, \ \mu = \mu_a\omega^a$ for easy reading of the equations. The superpotential $W$ can now 
be expressed as %and its 
\begin{eqnarray}
W & = & X_1 + i Y_1 
\end{eqnarray}
with 
\begin{eqnarray}
X_1 &=& q_0 - \frac{2D}{(p^0)^2} + 3 \frac{D_a\omega^a}{p^0} + 3 p^0\nu_a\omega^a - p^0 \mu \cr
Y_1 & = & - p^0 \nu - \frac{3 D_ax_2^a}{p^0} + 3 p^0 \mu_a x_2^a \label{X1Y1}
\end{eqnarray}
The covariant derivative of the superpotential $\nabla_aW$ is given by
%\begin{eqnarray}
%\nabla_a W  & =  & \frac{3}{2} \big( Y_{2a} + i X_{2a} \big) 
%\end{eqnarray}
%Here we have separated the real and imaginary parts of the covariant derivative by introducing the following quantities:
%\begin{eqnarray*}
%Y_{2a} &=& - \frac{2D_a}{p^0} + 2 p^0 \mu_a - 2 p^0 \nu_a - \frac{\nu_a}{\nu} Y_1 \\
%X_{2a} &=& - 4 p^0 \nu_{ab}\omega^b + \frac{\nu_a}{\nu} X_1
%\end{eqnarray*}

\begin{eqnarray}
\nabla_a W  = \frac{3}{2}\left(\Big( - \frac{2D_a}{p^0} + 2 p^0 \mu_a - 2 p^0 \nu_a - \frac{\nu_a}{\nu} Y_1 \Big)
 + i \Big(- 4 p^0 \nu_{ab}\omega^b + \frac{\nu_a}{\nu} X_1\Big)\right)
 \end{eqnarray}

 The supersymmetric solutions are obtained by setting the real and imaginary parts of $\nabla_aW$ to zero. The most 
 general solution to these equations
 are given by eqs.\eqref{eq:g2} and \eqref{eq:g1}. For non-supersymmetric configurations we also need to compute 
 $ \nabla_a\nabla_b W$. The real and imaginary parts of the above quantity are given respectively by %Setting 
 %\begin{equation}
% \nabla_a\nabla_b W =  f_{1ab} + i f_{2ab} \ ,
 %\end{equation}
% we find
 \begin{eqnarray*}
\frac{3}{2\nu} \Big(\nu_{ab} - 3\frac{\nu_a\nu_b}{\nu}\Big) X_1 + \frac{9p^0}{\nu}\big(\nu_a\nu_{bc} 
+ \nu_b\nu_{ac}\big) \omega^c - 6 p^0 \mu_{ab} 
\end{eqnarray*}
and 
\begin{eqnarray*}
 6p^0\nu_{ab} + \frac{3}{2\nu} \Big(\nu_{ab} - 3\frac{\nu_a\nu_b}{\nu}\Big) Y_1 
- \frac{9}{2\nu}\left( \frac{1}{p^0}\big(\nu_aD_b +\nu_bD_a\big)
+ 2p^0 \nu_a\nu_b - p^0 \big(\mu_a\nu_b + \mu_b\nu_a\big)\right)
\end{eqnarray*}
 We also need the inverse of the moduli space metric and its derivative:
 \begin{eqnarray*}
 g^{b\bar c} &=& -\frac{2\nu}{3} \left(\nu^{bc} - \frac{3}{\nu} x_2^bx_2^c\right) \\
 \partial_a g^{b\bar c} &=& i \left( \nu_a\nu^{bc} \frac{\nu}{3} D_{ade}\nu^{bd}\nu^{ce} - \delta_a^bx_2^c - \delta_a^cx_2^b\right)
 \end{eqnarray*}

We now substitute the above expressions in the equations of motion \eqref{extpot}. After a straightforward, but tedious computation
we obtain:
\begin{eqnarray}\label{realprt}
3 (p^0)^2 \nu_a\nu_b\omega^b  
+ \nu_{ab}\omega^b p^0 \big(Y_1 - p^0 \nu\big) 
&+& \big(X_1 - 3 p^0 \nu_b\omega^b\big) \left(\nicefrac{D_a}{p^0} - p^0 \mu_a\right)
\cr &+& p^0 \nu \nu^{bc}\mu_{ab} \left(\nicefrac{D_c}{p^0} - p^0 \mu_c\right) = 0 \ ,
\end{eqnarray}
for the real part of the equations of motion, and 
\begin{eqnarray}\label{imgnprt}
 \nu_a \big( 2 X_1^2 - 12 p^0 X_1 \nu_c\omega^c + 36 (p^0)^2 ( \nu_c\omega^c)^2 + Y_1^2 - (p^0)^2\nu^2 + 2 p^0 Y_1\nu\big) 
+  2 Y_1 \nu \left(\nicefrac{D_a}{p^0} - p^0 \mu_a \right) - \cr
  \nu^2 \nu^{bd}\nu^{ce} D_{ade}
 \left(\nicefrac{D_b}{p^0} - p^0 \mu_b\right) \left(\nicefrac{D_c}{p^0} - p^0 \mu_c\right)% + 2 (p^0)^2 \nu^2\mu_a
- 24 (p^0)^2 \nu \nu_{ab}\omega^b  \nu_c\omega^c  + 2 \nu^2\big(D_a + (p^0)^2 \mu_a\big)  =0 . \ 
\end{eqnarray}
 for the imaginary part. This is the most general form of the equation of motion for non-supersymmetric. Because of the 
complicated structure it is extremely hard to obtain the most general solution for the above equations. However, taking 
the clue from the existing solutions for their supersymmetric counter part we can look for appropriate ansatz to construct 
a class of new non-supersymmetric solutions for the above equations. We set 
\begin{equation}\label{ansz} 
x_2^a = {I^a}_b p^b x\ {\rm and}\ \omega^a = {I^a}_b p^b\omega \ ,
\end{equation}
 where the involution ${I^a}_b$ is assumed to satisfy \eqref{iabda}.{\footnote{Note that. it is not possible to redefine the 
 charges $p^a$ to get rid of the ${I^a}_b$ dependence because of the shift involved in defining $\omega^a$.}} Substituting 
 the above in eqs.\eqref{realprt} and 
 \eqref{imgnprt} we find, after a bit simplification:
 \begin{eqnarray}
 &&2 (p^0)^3 \chi x^4\omega + (1 - (p^0)^2\omega^2) (X_1 - 2 p^0 \chi x^2 \omega) + (p^0)^2 x\ \omega Y_1 = 0 \label{eom1x} \\
 &&2 x^2 X_1^2 - 12 p^0 X_1 \chi x^4 \omega + 12 (p^0)^2 \chi^2 x^6 \omega^2 + x^2 Y_1^2 - (p^0)^2 \chi^2 x^8 
 + 2 p^0 \chi Y_1 x^5 + 2 \chi^2 x^6  \cr &&  + 2 (p^0)^2 \chi^2 x^6 \omega^2 
 + \big(\nicefrac{2Y_1}{p^0}\big) \chi x^3 (1 - (p^0)^2 \omega^2) - \big(\nicefrac{\chi}{p^0}\big)^2 x^4 (1 - (p^0)^2\omega^2)^2 = 0 \label{eom2x}
 \end{eqnarray}
 We reproduce the expressions for $X_1$ and $Y_1$ after substituting the ansatz \eqref{ansz} in \eqref{X1Y1}:
 \begin{eqnarray*}
 X_1 &=& q_0 - \left(\nicefrac{2D}{(p^0)^2}\right) + \left(\nicefrac{3\chi}{p^0}\right)\omega 
 + 3 \chi p^0 \omega x^2 - p^0 D \omega^3 \cr
 Y_1 &=& - \left(\nicefrac{3\chi}{p^0}\right) x \big(1 - (p^0\omega)^2\big) - p^0\chi x^3 \ .
 \end{eqnarray*}
 Here for easy reading of the equations we have defined $\chi = D_a {I^a}_bp^b$. This gives a considerable simplification
 as we need to solve them only for the variables $x$ and $\omega$ in terms of the quantities $p^0,q_0,D$ and $\chi$. 
These equations can further be simplified by noting that they contain only even powers of $x$. Setting $x^2 = y$ and 
eliminating $y$ and $\omega$ respectively we find the following factorized form:
 \begin{equation}\label{eqw}
 f_1(\omega) f_3(\omega) F_3(\omega) = 0\ ,
 \end{equation}
 and 
 \begin{equation}\label{eqy}
 g_1(y) g_3(y) G_3(y) = 0 \ .
 \end{equation}
 Here $f_k(\omega), g_k(y)$ are polynomials of degree $k$ with respect to their arguments. Their explicit expressions 
 are given by
 \begin{eqnarray*}
 f_1(\omega) &=& 2\chi p^0 \omega + (p^0)^2 q_0 - 2 D \ , \cr
 g_1(y) &=& 4 \chi^2 (p^0)^2 y - \hat s \ ,
 \end{eqnarray*}
 and
 \begin{eqnarray*}
 f_3(\omega) &=& (2\chi^2 + \hat s) (p^0)^3 \omega^3 - 3 \chi (p^0)^2 \big(2 D - (p^0)^2 q_0\big)\omega^2
 + 6 \chi^2 p^0 \omega - \chi  \big(2 D - (p^0)^2 q_0\big) \cr
 g_3(y) &=& \chi^2(p^0)^6 \big(2\chi^2 + \hat s\big)^2 y^3 - 9 \chi^4 (p^0)^4 \hat s y^2 
 + 6 \chi^2 (p^0)^2 (\hat s)^2 y - (\hat s)^3
 \end{eqnarray*}
 with $\hat s =  \big(2 D - (p^0)^2 q_0\big)^2 -  4 \chi^2$. Solving the linear equations $f_1(\omega)=0=g_1(y)$
 gives rise to the supersymmetric attractors described in \cite{Manda:2015zoa}. We will now focus on the cubic 
 polynomials. The discriminants of $f_3(\omega)$ and $g_3(y)$ are given by $- 27 \chi^2 (\hat s)^3 (p^0)^6$
 and $ - 27 \chi^4(p^0)^{12} (\hat s)^7 (2 \chi^2 + \hat s)^2 \big(2 D - (p^0)^2 q_0\big)^2$ respectively. Both
 become negative for $\hat s>0$ and hence $f_3(\omega)=0=g_3(y)$ admit unique real valued solutions for 
 $\omega$ and $y$.  It is straightforward to verify that the resulting $\omega,y$ indeed provide a non-susy
 solution for the equations of motion \eqref{eom1x},\eqref{eom2x}. Further, it can be verified that the cubic
 polynomials $F_3(\omega)$ and $G_3(y)$ do not provide any solution for the equations of motion.

 To express the non-supersymmetric solution orderly in a closed form we will make the following rescaling of the 
 variables:
 \begin{equation}\label{rescale} 
  \omega\rightarrow \nicefrac{\tilde\omega}{p^0},\ y\rightarrow \nicefrac{\tilde y}{(p^0)^2},\ q_0 \rightarrow 
 \nicefrac{(\tilde q\chi + 2 D)}{(p^0)^2} \ .  
 \end{equation}
 The equations $f_3(\omega)=0=g_3(y)$ now take the simple form
 \begin{eqnarray*}
 (\tilde q^2 - 2) \tilde\omega^3 + 3 \tilde q\tilde\omega^2 + 6 \tilde\omega + \tilde q &=& 0 \cr
 (\tilde q^2 - 2)^2 \tilde y^3 - 9 (\tilde q^2 - 4) \tilde y^2 + 6  (\tilde q^2 - 4)^2 \tilde y - (\tilde q^2 - 4)^3 &=& 0 
 \end{eqnarray*}
 and the corresponding attractor solution is given by
 \begin{eqnarray}
 \tilde\omega &=& \frac{ f_-(\tilde q) - f_+(\tilde q) - 2^{1/3}\tilde q}{2^{1/3}(\tilde q^2-2)} \ ,  \\
 \tilde y &=& \frac{g_+(\tilde q) - g_-(\tilde q) + 2^{1/3} 3 (\tilde q^2 - 4)}{2^{1/3}(\tilde q^2 - 2)^2} \ ,
 \end{eqnarray}
 with
 \begin{eqnarray*}
 f_\pm(\tilde q) &=&  \big((\tilde q^2 - 2) (\tilde q^2 -4)^{3/2} \pm  \tilde q  (\tilde q^2 -4)^2 \big)^{1/3}  \ , \\
 g_\pm(\tilde q) &=& (\tilde q^2 - 4) \left(\tilde q (\tilde q^2 - 2)^3 \sqrt{\tilde q^2 - 4} \pm (\tilde q^8 - 8 \tilde q^6 + 6 \tilde q^4 
 + 40 \tilde q^2 - 2)\right)^{1/3} \ .
 \end{eqnarray*}
 The non-supersymmetric attractors can now be constructed from the above using \eqref{ansz} and the rescaling \eqref{rescale}.
 
 \section{Conclusion} 
 
 In this paper we have studied non-supersymmetric attractors in four dimensional $N=2$ supergravity coupled to $n$ vector
 multiplets with the purely cubic pre-potential. We have expressed the most general form of the equations of motion in terms 
 of a set of convenient variables involving the moduli fields ($\omega^a$ and $x_2^a$). We have used a generalized ansatz
 involving a constrained involutory matrix to solve the equations of motion. This gives rise to new branches of non-supersymmetric 
 attractors for every consistent choice of involutions. 
 
 It was possible to obtain an exact analytic expression for the solution because of the factorization in \eqref{eqw} and \eqref{eqy}.
 It would be interesting to see if it is possible to obtain an analogous expression without assuming any ansatz for the moduli. 
 This will help in classifying all non-supersymmetric attractors in these type of supergravity theories. It would also be interesting 
 to consider the flow equations and obtain generalized attractor equations to solve them. A first step towards this would be to 
 construct a fake superpotential for these non-supersymmetric attractors. Incorporating stringy corrections to the pre-potential
 too gives rise to rich structures. An issue of greater import is to look into the microscopic description of these new branches of 
 attractors in both supersymmetric as well as non-supersymmetric cases. Localization proves to be a powerful technique to 
 obtain the exact partition function which captures sub-leading corrections to the entropy. It is worth exploring whether it can 
 be used to understand the origin of these new branches in $N=2$ theories.

\end{document}